\documentstyle[11pt]{article}
\begin{document}
preprint HLRZ 38/93
\begin{flushright}
April, 1993\\
\end{flushright}
\begin{center}
\begin{Large}
Boundary-induced anisotropy of the avalanches
in the sandpile automaton\\
\end{Large}
\vspace{11pt}
Gongwen Peng\footnote{Address after
August 31, 1993: HLRZ, KFA J\"ulich, Postfach 1913, 5170
J\"ulich, Germany}\\
Institute of Physics, Chinese Academy of Sciences, P. O. Box 603,
Beijing 1000080, China\\
\vspace{7pt}
Hans J Herrmann\\
HLRZ, KFA J\"ulich, Postfach 1913, 5170 J\"ulich, Germany\\
\end{center}
\begin{abstract}
We study numerically  the avalanches in a
two--dimensional critical height sandpile model
with sand grains added at the center of the system. Smaller
avalanches near the center of the system are
isotropic. Larger avalanches are, however, affected
by the boundary of the system, to a degree that increases with
the avalanche size. Up to linear system size $L=1001$, we did not
find an obvious indication for lattice--induced anisotropy.
\end{abstract}
\newpage
\indent
Self--organized criticality (SOC) was originally proposed by Bak, Tang
and Wiesenfeld (BTW) [1] to explain the power--law correlations in
nature, e.g., the occurrence of fractal structures in space
and $1/f$ noise in time. Lattice automaton models displaying this
self--organized critical behavior have been investigated numerically by
several authors [2--6]. The most extensively studied models are
the sandpile models which were originally designed to capture some
features of the dynamics of avalanches on a sand pile. Despite the fact
that experiments on large real sand piles
found no SOC behavior, the theoretical
interest in these models remains high. This is because  our present
understanding of the dynamical systems with many degrees of freedom is
poor and in fact, the applications of the sandpile models are not
limited to real sand piles. As first proposed by BTW [1], one of the
characteristics of SOC is the self--similarity of the avalanches.
This is usually only true for the distribution of sizes, but not
for the individual avalanches, i.e., the internal
structures are not necessarily fractal. The avalanches in
the sandpile models are observed to be compact, to some degree similar
to Eden clusters [3,7].\\

In this paper we are interested in the spatial
distribution of avalanches in sandpile
models. Some avalanches reach the size of the system while
others stop growing  even if they do not touch the system boundary.
In the following we will see that the shape of the system boundary has
a long--range effect on the spatial distribution of avalanches,
i.e., even the avalanches
whose surfaces are far away from the system's boundary are
influenced by the boundary shape. Since this boundary--induced effect
scales with the system size, it can not be considered
to be a ``finite--size effect''. Systems with anisotropic boundaries will
have anisotropic avalanches of larger sizes and isotropic avalanches of
smaller sizes. This boundary-anisotropy
is different from the lattice--induced
anisotropy which has been observed in some sandpile lattice automata [8]
and which we did not observe even
considering systems of sizes up to $L=1001$.
Cluster anisotropy due to anisotropic lattices has also been observed in
simulations of irreversible growth models such as DLA [9] and the
Eden model [10].\\

We carry out the simulation of the critical height model
on a two--dimensional square  lattice
where a sand column topples when its height exceeds
a threshold. Sand grains are added to  the  system one by one at the
central site of the lattice (which we  mark as the origin). Specifically,
we consider  an  $L\times  L$  square  lattice  with  open  boundary
condition ($L$ is chosen to be an odd integer so that there is  a  unique
central site). The dynamics of the system proceeds by two kinds of
operations. First, if the state configuration is stable,  a  sand
grain is added at the origin so that the height $h$ of the sand column at
that site is increased by one unit:
\begin{equation}
h(0,0)\rightarrow h(0,0)+1
\end{equation}
Second, if at any site the height of the sand column is greater than $4$,
\begin{equation}
h(i,j)>4
\end{equation}
then this site is unstable and topples in the following way:
\begin{equation}
h(i,j)\rightarrow h(i,j)-4
\end{equation}
\begin{equation}
h(i,j\pm 1)\rightarrow h(i,j\pm 1)+1
\end{equation}
\begin{equation}
h(i\pm 1,j)\rightarrow h(i\pm 1,j)+1
\end{equation}
When a toppling occurs at a boundary site, sand grains which go out of
the system never come back. Although this automaton is deterministic,
it exhibits self--organized critical behavior [11], characterized by
power--law distributions of the avalanche sizes and lifetimes.\\

After a transient period of adding sand grains, the system reaches
the SOC state where the average inflow is equal to the average outflow.
It is checked that the initial configuration is irrelevant to the final
results. Usually we start from an initial configuration where at each
site the sand column has a height $h$ being a random integer between $1$
and $4$.\\

An avalanche is defined as the
sites that have toppled at least once during a chain-reaction
of topplings. To get  better statistics for the avalanches, we
record for each site the number $T$ of topplings for many  avalanches.
There exist two ways of recording: for a site which topples $m$ times
during an avalanche, we increase its $T$ either by $m$ or by $1$.
Let us focus on the number $T$ of
topplings on the sites located on the four
sides of a square of length $l=81$. The center of this square is the
origin and the system size is $L=99$. In fig.1 we plot the normalized
number $P$ of topplings ($T$  divided by the sum of $T$ for all sites)
against the polar angle. The two curves in fig.~1 correspond to the two
ways of recording which yield close  results and we see that
there is a small difference between these two ways.
A systematic comparison between the two types of toppling has
been made in ref.~[3].\\

To see whether the avalanches are isotropic
or not, it is better to compare the number of topplings for sites which
are located on a circle. Sites whose distances to the origin $r$ satisfy
$r_c - \frac{1}{2}\leq r < r_c + \frac{1}{2}$   are
identified to be on a circle of radius $r_c$. Fig.2(a) shows the plot
of the normalized number of topplings  for sites located on a circle of
diameter $d=81$ against their polar angle $\theta $,
and fig.2(b) is a polar
representation for this angular distribution. It is clear that the
avalanches are anisotropic since they grow more often in the diagonal
directions than in the axial ones. Fig.3 shows the contour
lines for the number of topplings in the first quadrant for two
different systems of sizes $L=99$ and $L=301$. One notices that the
avalanches are isotropic for smaller sizes but anisotropic for
the larger ones. Interestingly, the amount of anisotropy only seems
to depend on the relative distance from the boundary implying
that the effect scales with the system size.\\

Supposing a contour line has a distance $r$ from the origin
in (1,0) direction and a distance $\bar{r}$ in (1,1) direction,
then $\delta =(\bar{r}-r)/r $ measures the degree of
anisotropy of avalanches.
Obviously, $\delta$ is a function of $r$. Let us consider the
number of topplings of the sites located on a circle of radius $r$.
If the avalanches are isotropic, the number $n$ of topplings in (1,0)
direction should be equal to the number $\bar{n}$ of topplings in
(1,1) direction. For anisotropic avalanches,
$\epsilon=(\bar{n}-n)/n$ provides another way of characterizing
the anisotropy. In fig.4(a) and fig.4(b) we show $\delta$
and $\epsilon$ for different lattice sizes.
It is astonishing to see how well the different system sizes
collapse on top of each other reinforcing our speculation that
the correct scaling factor is the linear system size $L$.\\

If the square boundary is replaced by a circular one, the
above--mentioned anisotropy seems to
disappear. Fig.5 shows the normalized
number of topplings for the sites located on a circle of diameter
$d=81$ and the system boundary being a circle of diameter $D=99$.
Besides a weak ``noisy'' structure,
there is no visible peak in fig.5(a). This indicates
that the anisotropy in figs.2 and 3 is caused by the
anisotropic boundary shape (boundary-induced).
One sees that the complicated ``noisy'' structure is
identical in each quadrant.
We believe that this structure just reflects
the fact that the surrounding boundary is not a perfect circle but a
discretized approximation of a circle on a square lattice.\\

In order to explore in more detail if there exists a lattice--induced
anisotropy like the one observed in DLA [8] and Eden [9] models,
we also considered very large systems with circular boundaries. Up to a
system size of diameter $D=1001$, we did not find any indication for
lattice--induced anisotropy. In fig.6 we recorded
the number of topplings according to the polar angles of the sites
irrespective of their distances to the origin (by dividing the system
into $80$ equal sectors). No obvious peak exists in fig.6 except for
the ``noisy'' structure
that was also observed for the smaller system sizes.
We conclude that the boundary shape
affects the avalanches of all sizes to a degree that depends
on the size of the avalanche.
The smaller the size, the weaker is the anisotropy.\\

It is still possible that the lattice anisotropy has
an effect on the avalanche shapes but that
this is difficult to observe for the two--dimensional critical--height
model with our present computer facilities. Recently Puhl [8] has given
many interesting results for the anisotropy of the critical--slope sand
pile model. He has observed anisotropy of the avalanches as well as the
piles due to the anisotropy of lattice. He also studied sandpiles
on a random lattice [12] and found agreement with experiments
on real small sandpiles [13].\\

In summary we have observed that the shape of the boundary of the
system has a long--range influence on the local distribution of
avalanches. For smaller avalanches this influence is negligible
while larger avalanches feel the boundary shape
over distances that increase linearly with the system
size. Lattice--induced anisotropy was not observed for systems
up to a linear size $L=1001$. More than ten days of CPU time on DEC 5200
have been used to obtain the results reported here.\\

G. W. P. was supported by the Institute of Physics, Chinese Academy of
Sciences, and partially by grant LWTZ--1298 from Chinese Academy
of Sciences. H.J.H. thanks the Academia Sinica for hospitality.\\

\newpage
\noindent
{\bf Figure captions:}\\
\indent

Fig.1: Normalized number $P$ of topplings for the sites located
on the four sides of a square of length $l=81$ against their
polar angle $\theta $. The system has a square open boundary and a size
$L=99$. \hspace{0.3cm} $2.3\times 10^6$ avalanches
were used. The two curves correspond to two ways of recording the
number of topplings.\\

Fig.2: (a) Normalized number $P$ of topplings for the sites located
on a circle of diameter $d=81$ against their polar angle $\theta $.
The system has a square open boundary with size $L=99$.
\hspace{0.3cm} $2.3\times 10^6$ avalanches were used.
(b) Polar representation for the
angular distribution in (a), i.e. the distance of a small circle
from the center is the value of $P(\theta)$ in the direction $\theta$.\\

Fig.3: Contour lines, i.e. sites of equal
number of topplings  of the sites
in the first quadrant of the systems with sizes (a) $L=99$, (b)
$L=301$. The number of avalanches used are (a) $1.9 \times 10^6$
and (b) $2.1 \times 10^5$. The smoother curves
are circular guides to the eyes.\\

Fig.4: Semi-log plot of $\delta$ (a) and $\epsilon$ (b) as a function
of $r/L$ for different lattice sizes $L=99$ (dotted), $L=201$
(dash--dot), $L=301$ (dash--double--dot) and L=401 (solid).
We observe an excellent data collapse.\\

Fig.5: (a) Normalized number $P$ of topplings for the sites located
on a circle of diameter $d=81$ against their polar angle $\theta $.
The system has a circular boundary of diameter of $D=99$.
\hspace{0.3cm} $2.1 \times 10^6$ avalanches were used.
(b) Polar representation for the angular distribution
in (a), i.e. the distance of a small circle  from the center is the value
of $P(\theta)$ in the direction of $\theta$.\\

Fig.6: Normalized total number $P$ of topplings for the sites within
a sector of opening angle $4.5^{\circ }$ as a function
of the polar angle $\theta $ for a
system with a circular boundary of diameter $D=1001$.
\hspace{0.3cm} $6.6 \times 10^4$ avalanches were used.

\newpage
\begin{thebibliography}{99}
\bibitem{1}  {P. Bak, C. Tang and K. Wiesenfeld, Phys.
Rev. Lett. {\bf 59} (1987) 381; Phys. Rev. A {\bf 38} (1988) 364.}
\bibitem{2} {L. P. Kadanoff, S. R. Nagel, L. Wu and S. M. Zhou, Phys. Rev.
A {\bf 39} (1989) 6524.}
\bibitem{3} {P. Grassberger and S. S. Manna, J. Phys. (Paris)
{\bf 51} (1990) 1077.}
\bibitem{4} {S. S. Manna, L. B. Kiss and J. Kert\'esz, J. Stat.
Phys. {\bf 61} (1990) 923.}
\bibitem{5} {E. N. Miranda and H.  J.  Herrmann,  Physica  A  {\bf 175}
(1991) 339.}
\bibitem{6} {S. S. Manna, Physica A {\bf 179} (1991) 249.}
\bibitem{7} {L. Pietronero, P. Tartaglia and Y. --C. Zhang,
Physica A {\bf 173} (1991) 22.}
\bibitem{8} {H. Puhl, Physica A {\bf 182} (1992) 295, and Ph. D thesis,
HLRZ, KFA J\"ulich, 1992.}
\bibitem{9} {P. Meakin in: Phase Transition and Critical
Phenomena, A. C. Domb and J. L. Lebowitz, eds. (Academic Press,
New York, 1988), p. 336.}
\bibitem{10} {D. Dhar, in: On Growth and Form, H. E. Stanley and
N. Ostrowsky, eds. (Martinus Nijhoff Publishers, Dordrecht,
1986), p.288.}
\bibitem{11} {K. Wiesenfeld, J. Theiler and B. McNamara, Phys.
Rev. Lett. {\bf 65} (1990) 949.}
\bibitem{12} {H. Puhl, Physica A, preprint HLRZ 16/93.}
\bibitem{13} {M. Bretz, J.B. Cunningham, P.L. Kurczynski and
F. Nori, Phys. Rev. Lett. {\bf 69} (1992) 2431.}
\end{thebibliography}
\end{document}